\documentclass[12pt, letterpaper]{JHEP3}

\usepackage{graphicx}
\usepackage{amssymb}

\title{Evaporation of large black holes in AdS: coupling to the evaporon}

\author{Jorge V. Rocha\\ Department of Physics, University of California Santa Barbara, CA 93106\\ E-mail: \email{jrocha@physics.ucsb.edu}}

\abstract{
Large black holes in an asymptotically AdS spacetime have a dual description in terms of approximately thermal states in the boundary CFT.  The reflecting boundary conditions of AdS prevent such black holes from evaporating completely.  On the other hand, the formulation of the information paradox becomes more stringent when a black hole is allowed to evaporate.  In order to address the information loss problem from the AdS/CFT perspective we then need the boundary to become partially absorptive.
We present a simple model that produces the necessary changes on the boundary by coupling a bulk scalar field to the evaporon, an external field propagating in one extra spatial dimension.  The interaction is localized at the boundary of AdS and leads to partial transmission into the additional space.  The transmission coefficient is computed in the planar limit and perturbatively in the coupling constant.  Evaporation of the large black hole corresponds to cooling down the CFT by transferring energy to an external sector.
}

\begin{document}

\section{Introduction}

In 1974 Hawking showed that black holes radiate~\cite{Hawking:1974sw}.
This phenomenon is a purely quantum mechanical effect and is viewed as a milestone of quantum gravitational systems.
Moreover, the spectrum of the radiation produced is that of a black body.
This led to the formulation of the famous information paradox~\cite{Hawking:1976ra} shortly after, since information sent into a black hole seems to be lost after it evaporates.
This remains an outstanding problem whose resolution would most likely lead us to a better understanding of quantum gravity.

In the past ten years we have acquired a tool, namely the AdS/CFT correspondence~\cite{Maldacena:1997re,Gubser:1998bc,Witten:1998qj}, which allows us to study gravitational systems in terms of a dual gauge theory.
The manifest unitarity of the field theory strongly suggests that no information loss occurs during the process of black hole evaporation~\cite{Lowe:1999pk} and one such proposal for how this may come about was formulated in~\cite{Maldacena:2001kr}.
Recently the information paradox has been discussed from the dual gauge theory point of view in the context of a matrix model~\cite{Iizuka:2008hg}.
However, we still lack a good description of the evaporation process from the gravity side.
In order to use the AdS/CFT duality we must consider black holes in asymptotically anti-de Sitter spacetimes.
In many aspects AdS behaves just like a box, the travel time for a null geodesic to cross the whole space being finite.
Of course, if we place a small enough black hole in AdS it will evaporate before it gets to feel this finiteness of the surrounding space.
Also, these small black holes do not have a direct interpretation in terms of the dual CFT as they are classically unstable~\cite{Witten:1998zw,Horowitz:1999jd}.
However, the so called large black holes which have a positive specific heat~\cite{Hawking:1982dh} (in contrast with the small black holes) do not evaporate completely and instead reach a configuration of thermal equilibrium with the surrounding gas of particles.
Arbitrarily massive black holes in AdS have arbitrarily high Hawking temperature but it has been argued that this does not lead to a gravitational instability~\cite{Hemming:2007yq}.
Such black holes have a dual description in terms of a finite temperature field theory~\cite{Witten:1998zw}.
One would think {\it a priori} that the class to which the black hole belongs is determined by the radius of its horizon, $r_H$, being greater or less than the characteristic scale $R$ of AdS set by the cosmological constant.
However, it has been shown~\cite{Horowitz:1999uv} that this transition from small to large black holes occurs at a value of the ratio $r_H/R$ which falls off with a negative power of the AdS radius in Planck units, and so can be parametrically small.

\FIGURE[t]{
 \includegraphics[width=25pc, bb= 0pt 0pt 430pt 130pt]{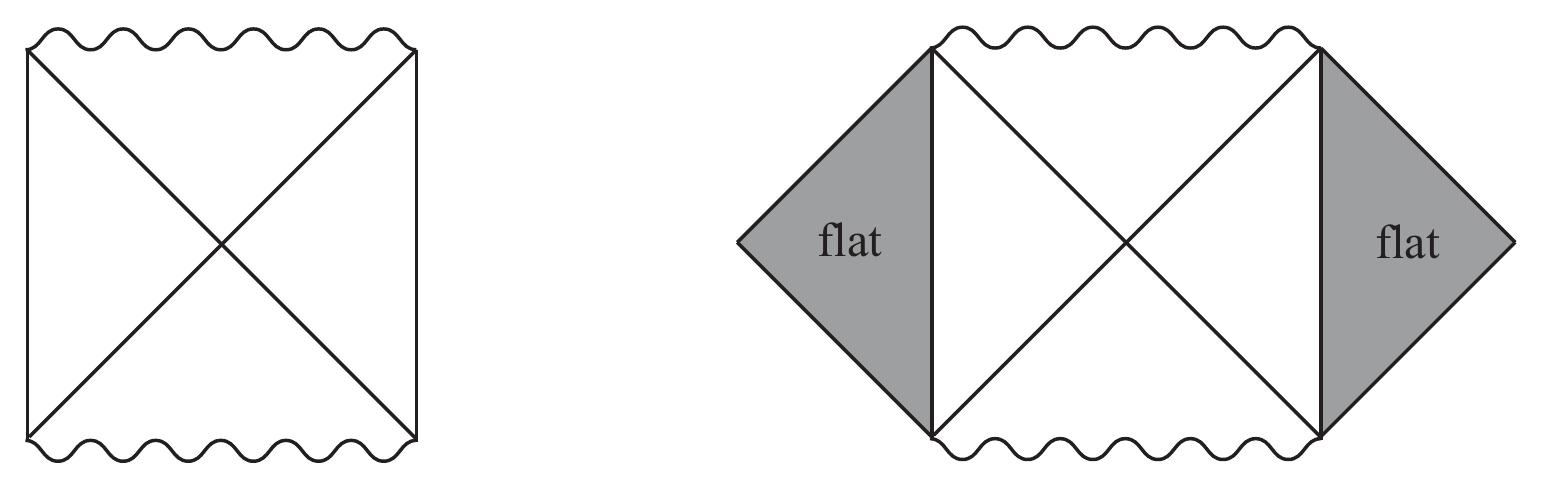}
 \caption{The Penrose diagram on the left shows the eternal Schwarzschild-AdS black hole.  In the diagram on the right (which is technically not a Penrose diagram) we attach a $1+1$ dimensional flat space to the boundary of AdS, resulting in a picture resembling the eternal black hole geometry in flat space.}
 \label{penrose}
}

The presence of a negative cosmological constant effectively imposes reflecting boundary conditions on the bulk fields and this prevents the large black holes from evaporating.
Clearly, this situation does not provide a good context to address the information paradox.
In this paper we attempt to bring the paradox back on stage by setting up a toy model that effectively changes the boundary conditions on AdS so that it is not totally reflecting.
In this framework a minimally coupled bulk scalar field in ${\rm AdS}_5$ is coupled to a scalar field propagating in an additional $(1+1)$-dimensional flat space.
The interaction is localized on the boundary of AdS and at the origin of the added flat space (see Figure~\ref{penrose}).
The purpose of this scheme is to allow part of the radiation from the black hole to be transmitted into the extra space, thus permitting the black hole to evaporate completely.
In principle, the rate of evaporation can be obtained by determining the spectrum of the Hawking radiation as in~\cite{Hawking:1974sw}, and then integrating it over all the modes~\footnote{A calculation of the Hawking radiation from AdS black holes has been done in~\cite{Hemming:2000as} but it only addresses the onset of black hole formation.}.
This calculation will not be addressed in the present work but is currently under investigation.

The description in terms of the dual gauge theory is as follows.
The AdS/CFT correspondence~\cite{Gubser:1998bc,Witten:1998qj} in its weakest form relates the CFT generating functional for a single trace operator ${\cal O}_\Delta$ to the type IIB supergravity partition function~\footnote{There are well known issues associated with the formulation of the AdS/CFT correspondence in Lorentzian signature~\cite{Balasubramanian:1998sn,Son:2002sd}.  We will employ analytic continuation to perform our calculations in Euclidean signature.}:
\begin{equation}
\left< e^{i \int_{\cal B} d^4z \, \sigma({\bf z}) {\cal O}_\Delta({\bf z})} \right>_{CFT}  =  \exp \left\{ i \left. {\rm extr} \, S_{SUGRA}[\Phi] \right|_{\Phi({\bf z},\varepsilon) = \varepsilon^{4-\Delta} \sigma({\bf z}) } \right\}     \ .
\end{equation}
The supergravity field $\Phi$ is dual to the operator ${\cal O}_\Delta$.
Consider now promoting the source $\sigma$ to a propagating field by giving it a kinetic term.
The path integral now runs over this extra field and we should then have
\begin{equation}
\int {\cal D}\sigma {\cal D}X  e^{i S_{CFT} + i \int_{\cal B} \sigma \, {\cal O}_\Delta + i S[\sigma] } 
=  \int {\cal D}\sigma  \exp \left\{ i \left. {\rm extr} \, S_{SUGRA}[\Phi] \right|_{\Phi({\bf z},\varepsilon) = \varepsilon^{4-\Delta} \sigma({\bf z}) } + i S[\sigma] \right\}     \ ,
\label{partfunc}
\end{equation}
where we are denoting the fields on the CFT side collectively by $X$.
On the right hand side, the supergravity action is evaluated on classical solutions that satisfy the given asymptotic form.
A natural way to impose this condition is to add an interaction term of the form
\begin{equation}
\int_{\cal B}  \varepsilon^{4-\Delta} \sigma \, \Phi  \ .
\end{equation}
This will be done in section~2.
Therefore, we see that the dual description of a large black hole in AdS evaporating under this particular deformation is that of a CFT cooling down by transferring energy from its fundamental fields to an external field.
For simplicity we take the scalar field $\sigma$ (which we shall call the evaporon from now on) to live in $1+1$ dimensions so that it only couples to the s-wave of the bulk field.

There is an extra complication which must be taken into account.
As it stands, the self-energy for the additional scalar field $\sigma$ diverges as we take the cutoff $\varepsilon$ to the boundary of AdS.
Thus, counterterms are needed to renormalize the theory.
These will be carefully computed in section~3.
Nevertheless, we can see their origin from the CFT side: the OPE of the operator ${\cal O}_\Delta$ with itself behaves like
\begin{equation}
{\cal O}_\Delta(z) {\cal O}_\Delta(z')  \sim  \frac{1}{(z-z')^{2\Delta}}    \ ,
\end{equation}
and so when two operators of the form $\sigma(\tau,x) {\cal O}_\Delta(\tau,\vec{z})$ approach each other, in general we will get divergences.
Integrating over the spatial coordinates $\vec{z}$ then leaves divergences up to order $\varepsilon^{4-2\Delta}$, where $\varepsilon^{-1}$ represents the cutoff on momentum.
If we take our bulk scalar field to be the dilaton the dual operator is~\cite{Klebanov:1997kc} ${\cal O}_\Phi \propto Tr F^2$, where $F_{\mu\nu}$ represents the $SU(N)$ field strength, and has conformal dimension $\Delta = 4$.
We then expect counterterms of the form $\varepsilon^{-4}\sigma^2(\tau,0)$.
Of course, less singular counterterms which involve higher derivatives of the evaporon will also be generated: for the case of the dilaton we will find counterterms in $\dot{\sigma}^2(\tau,0)$ and $\ddot{\sigma}^2(\tau,0)$, corresponding to quadratic and logarithmic divergences, respectively.
The field $\sigma$ is dimensionless and in principle we could have non-quadratic (in $\sigma$) countertems.
However, it is easy to see from the gravity side that these are absent because they can only be generated from interactions in the bulk and these are suppressed in $1/N$~\cite{Maldacena:1997re}.
Therefore, these are the only possible counterterms.
The other possible combinations are either total derivatives or are related to the above ones by integration by parts.

If we take the bulk field $\Phi$ to have a mass that saturates the Breitenlohner-Freedman bound~\cite{Breitenlohner:1982bm}, $m^2 R^2 = -4$, the dual operator ${\cal O}_\Delta$ will have conformal dimension $\Delta = 2$.
Hence, this case is simpler, in the sense that we only need one counterterm to cancel the logarithmic divergence.
We will consider this special case in our calculations as well but our real interest lies with the dilaton as massless fields easily lend themselves to a geometric optics treatment for the calculation of the Hawking radiation from black holes that we would ultimately like to place in AdS.

Finally, we note that we will be working in an extremely simplified limit of the full supergravity by just considering the one bulk scalar field isolated from the rest of the bulk modes.
However, this is a good approximation in the large $N$ limit when all the fields decouple.

The outline of the paper is as follows.
In section~2 we set up the model that will be adopted for the remaining of the paper on more concrete grounds.
Section~3 deals with the computation of the counterterms, which will be carried out explicitly for the case $\Delta = 2$ first and then for $\Delta = 4$.
These counterterms are in fact crucial for our purposes and in section~5 we will see explicitly that without their inclusion we would not be able to obtain a finite transmission coefficient.
We devote section~4 to finding the (dis)continuity conditions on the interaction interface between the bulk field and the evaporon.
These conditions supplement the equations of motion and are needed to solve the scattering problem which reduces to a set of two (coupled) equations which describe wave scattering by a delta function potential in one dimension.
This computation is done in section~5 where we obtain a result for the transmission coefficient of the interface.
Section~6 contains the conclusions and some discussion.

\section{The setup}

We consider 5-dimensional Anti-de Sitter space (${\rm AdS}_5$) with curvature $R$ for which the metric, in global coordinates $(t,r,{\bf \Omega_3})$, reads:
\begin{equation}
ds^2 = - f(r) dt^2 + f^{-1}(r) dr^2 + r^2 d\Omega_3^2  \ ,
\end{equation}
where
\begin{equation}
f(r) =  1 + \frac{r^2}{R^2}  \ .
\end{equation}
With this parametrization the boundary has topology  ${\mathbb R}\times S^3$.
Ultimately we would like to put a black hole in this space.
For the simplest case of an AdS-Schwarzschild solution this amounts to adding a term $-\frac{r_0^2}{r^2}$ to the function $f(r)$.
However, in this paper we will be concerned only with the asymptotic region and so we consider the metric as given above.

In fact, for reasons of computational simplicity the calculations will be performed not in global coordinates, but rather in Poincar\'{e} coordinates, for which the AdS metric takes the following form:
\begin{equation}
ds^2 = \frac{R^2}{y^2} \left[-d\tau^2 + d\vec{z}^2 + dy^2 \right] \ .
\end{equation}
The patch defined by $\mathbb{H} \equiv \left\{ (\tau,\vec{z},y) | y \geq 0 \right\}$ covers half of ${\rm AdS}_5$ and the boundary is identified with $y=0$.
Therefore, the boundary in these coordinates has topology $\mathbb{R} \times {\mathbb R}^3$.
Except for this difference, the second metric is the large $r$ limit of the first one and can be obtained by the transformation $r = R^2/y$.
Nevertheless, since we are interested in black holes in (global) AdS we mimic the finite volume of the $S^3$ by periodically identifying the spatial coordinates $\vec{z}$.
Therefore, integrals over the full space yield a volume $V_3$.
Since we are replacing the $S^3$ by a 3-torus the black holes we can consider in this model are actually black branes and these always fall in the category of `large' black holes.

In order to allow part of the radiation from the black hole to leak out of AdS instead of totally reflecting we take the bulk scalar field $\Phi(\tau,\vec{z},y)$ to couple to a scalar field $\sigma(\tau,x)$, the evaporon, which propagates in $\mathbb{R} \times \mathbb{R}^+$.
Even though we are generalizing $\sigma$ to depend on the extra coordinate $x$, the argument around equation~(\ref{partfunc}) still follows through unchanged.
Since the evaporon lives on a real half-line we should supplement it with some boundary condition.
Ultimately we want to allow energy to be transfered between the bulk field and the external field so we implement this by extending $\sigma$ to the full real line and require it to be an even function of $x$.

In the context of AdS/CFT, calculations of 2-point functions on the boundary usually require a careful regularization which amounts to introducing a cutoff in the bulk geometry at $y = \varepsilon$ and then finding the solutions to the equations of motion subject to the Dirichlet boundary condition  $\Phi(\tau,\vec{z},\varepsilon) = \varepsilon^{4-\Delta} \overline{\Phi}(\tau,\vec{z})$.
In view of this we will take the two fields to interact only along the hypersurface ${\cal S} \equiv \left\{ (\tau,\vec{z},y) | y = \varepsilon \right\}$ and in the end we want to push this to the boundary of AdS, $\varepsilon \rightarrow 0$.

Therefore, we shall take the following action as our starting point:
\begin{eqnarray}
S[\Phi,\sigma] & = & S_\Phi[\Phi] + S_\sigma[\sigma] + S_{int}[\Phi,\sigma] + S_{c.t.}[\sigma]  \ ,  \label{action} 
\\
\nonumber \\
S_\Phi[\Phi] & = &  - \frac{1}{2\kappa^2} \int_\mathbb{H} d\tau d^3z dy \, \sqrt{-g} \left[ \frac{1}{2} g^{ab} \partial_a \Phi \partial_b \Phi + \frac{1}{2} m^2 \Phi^2 \right]     \ , \nonumber \\
S_\sigma[\sigma] & = &  - \int_{\mathbb{R} \times \mathbb{R}^+} d\tau dx \, \frac{1}{2} \left[ -(\partial_\tau \sigma)^2 +(\partial_x \sigma)^2 \right]     \, \nonumber \\
S_{int}[\Phi,\sigma] & = &  \lambda \int_{\cal S} d\tau d^3z \, \sqrt{-h} \, \varepsilon^{4-\Delta} \, \Phi(\tau,\vec{z},\varepsilon) \, \sigma(\tau,0)    \ ,  \nonumber
\end{eqnarray}
where $g_{ab}$ denotes the bulk metric, $g$ is its determinant and $h$ represents the determinant of the metric induced on the hypersurface ${\cal S}$.
The conformal dimension of the operator ${\cal O}_\Delta$ dual to the bulk scalar field is related to the mass $m$ through~\cite{Gubser:1998bc,Witten:1998qj} 
\begin{equation}
\Delta = 2 + \sqrt{4 + R^2 m^2}   \ .
\end{equation}
The constant $\kappa^2$, which essentially comes from dimensionally reducing on the $S^5$, has mass dimension $-3$ so the bulk field is dimensionless, as well as the evaporon living in a $1+1$ Minkowski space.
Therefore, the coupling constant $\lambda$ has mass dimension $8-\Delta$.
The particular way in which the two fields are coupled forces the evaporon to be a real scalar field.
Note also that the field redefinition  $\Phi = \sqrt{2}\kappa \Phi^{\rm new}$  brings the normalization of the kinetic term for $\Phi$ into canonic form at the expense of rescaling the coupling constant, $\lambda^{\rm new} = \sqrt{2}\kappa \lambda$.
The piece containing the counterterms, $S_{c.t.}$, will be determined in the following section.

\section{Computation of the counterterms}

\subsection{The tachyonic scalar field case ($\Delta=2$)}

As stated in the introduction we are mainly interested in the case of a massless scalar field in the bulk but we will first consider the case of a massive scalar saturating the Breitenlohner-Freedman bound~\cite{Breitenlohner:1982bm}, which corresponds to the minimal addition of counterterms for the evaporon.
In this situation we only expect logarithmic divergences in $\varepsilon$.

\FIGURE[t]{
 \includegraphics[width=22pc, bb= 0pt 0pt 280pt 35pt]{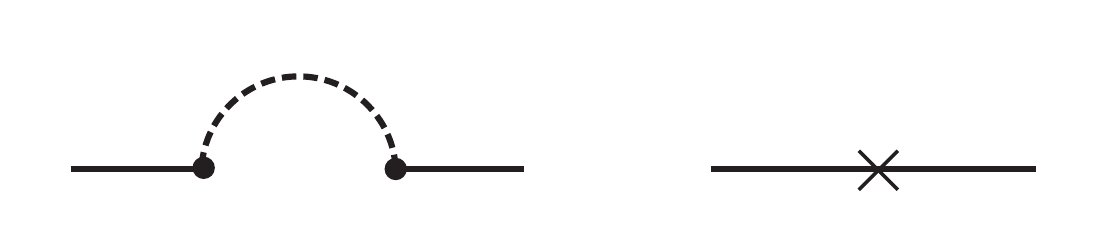}
 \caption{The two Feynman diagrams contributing to the evaporon (solid line) self-energy to order $O(\lambda^2)$.  The dashed line represents the dilaton propagating in the bulk of AdS.  The counterterms on the right are required to cancel the divergences arising from the diagram on the left.}
 \label{counterterms}
}

The counterterms needed can be found by carefully subtracting the divergences in perturbation theory.
At order $O(\lambda^2)$ there is a correction to the free $\sigma$-propagator which comes from a dilaton mediating the propagation of the evaporon.
The corresponding Feynman diagram is shown in Figure~\ref{counterterms}.
Each vertex contributes a factor of $i\sqrt{2}\kappa\lambda \frac{R^4}{\varepsilon^2}$.
The massive scalar bulk-to-bulk propagator between two points on the cutoff hypersurface ${\cal S}$ is expressed in terms of a hypergeometric function by~\cite{Burgess:1984ti}
\begin{equation}
\frac{1}{i} G_\Phi(\tau-\tau',\vec{z}-\vec{z}') = \frac{1}{8\pi^2 R^3} \xi^2 \,_2F_1 (1,3/2 ; 1; \xi^2)  
= \frac{1}{8\pi^2 R^3} \frac{\xi^2}{(1-\xi^2)^{3/2}}   \ ,
\end{equation}
where $\xi$ is given by
\begin{equation}
\xi = \frac{2\varepsilon^2}{2\varepsilon^2 - (\tau-\tau')^2 + (\vec{z}-\vec{z}')^2}  \ .
\label{xi}
\end{equation}
In the Lorentzian formulation care must be taken in the choice of sheet due to the branch cut extending from $1$ to $\infty$.
This becomes important for timelike separated points on the boundary but can be circumvented by Wick rotating, $\tau \rightarrow -i\tau_E$, and performing the calculation in Euclidean signature where no branch cuts are ever encountered, since $\xi \leq 1$ in this case.

The next step would be to Fourier transform the dilaton propagator to momentum space and obtain $\widetilde{G}_\Phi(\omega,\vec{k})$.
However, since $\Phi$ only interacts with a field that does not propagate on AdS, momentum conservation at the vertices implies that all we need is $\widetilde{G}_\Phi(\omega,\vec{0})$.
Hence,
\begin{equation}
\frac{1}{i} \widetilde{G}_\Phi(\omega,\vec{0}) 
=  \int d\tau d^3z \, e^{i \omega \tau} \frac{1}{i} G_\Phi(\tau,\vec{z})
=  -\frac{i}{8\pi^2 R^3} \int d\tau_E d^3z \, e^{\omega \tau_E}  \frac{\xi^2}{(1-\xi^2)^{3/2}}      \ .
\end{equation}
Under the following coordinate transformation
\begin{eqnarray}
\tau_E &=& \rho \sin \theta  \ , \nonumber \\
|\vec{z}| &=& \rho \cos \theta  \ ,
\label{change}
\end{eqnarray}
with $-\frac{\pi}{2} \leq \theta \leq \frac{\pi}{2}$, the above expression becomes
\begin{equation}
\widetilde{G}_\Phi(\omega,\vec{0}) 
=   \frac{1}{8\pi^2 R^3} {\rm Vol}(S^2) 
    \int_{-\pi/2}^{\pi/2}  d\theta \cos^2(\theta)  \int_0^\mu d\rho  \, \rho^3 
    \frac{\xi(\rho)^2}{(1-\xi(\rho)^2)^{3/2}}   +  \dots     \ .
\end{equation}
The dots contain terms that do not contribute to the self-energy of the evaporon after we take the limit $\varepsilon \rightarrow 0$.
We have introduced an infra-red cutoff~\footnote{This is related to the fact that we are working in Poincar\'e coordinates.  If we had chosen to work in global coordinates the boundary of AdS would have topology $\mathbb{R} \times S^3$ and the IR divergence would have been absent.} at $\rho = \mu$.
Changing integration variable from $\rho$ to $\xi$, performing the integral and Taylor expanding around $\varepsilon = 0$ we obtain
\begin{equation}
\widetilde{G}_\Phi(\omega,\vec{0})  =  - \frac{\varepsilon^4}{2R^3} \ln\left(\frac{\varepsilon^2}{\mu^2}\right)  +  O(\varepsilon^5)    \ .
\end{equation}

Therefore, the contribution to the evaporon self-energy from the first diagram in Figure~\ref{counterterms} is equal to
\begin{equation}
\left( i \sqrt{2} \kappa \lambda \frac{R^4}{\varepsilon^2} \right)^2 V_3 \frac{1}{i} \widetilde{G}_\Phi(\omega,\vec{0}) 
=  - i \kappa^2 \lambda^2 R^5 V_3  \ln\left(\frac{\varepsilon^2}{\mu^2}\right)  +  O(\varepsilon)      \ .
\end{equation}
Cancellation of the divergences then uniquely specifies the counterterms:
\begin{equation}
S_{c.t.}[\sigma] =  \frac{1}{2} \kappa^2 \lambda^2 R^5 V_3  \,  \ln\left(\frac{\varepsilon^2}{\mu^2}\right)
                    \int d\tau dx \, \delta(x) \sigma^2        \ .
\label{counter1}
\end{equation}
Lorentz symmetry is broken by the interaction between the two fields and as a consequence the counterterms are not Lorentz invariant.

\subsection{The massless scalar field case ($\Delta=4$)}

Now we turn to the more interesting case of the dilaton.
As we discussed in the introduction we expect quartic, quadratic and logarithmic divergences in the evaporon effective theory as we take $\varepsilon \rightarrow 0$.
The counterterms needed can be computed from the same Feynman diagram in Figure~\ref{counterterms}.
However, now each vertex contributes a factor of $i\sqrt{2}\kappa\lambda \frac{R^4}{\varepsilon^4}$ and the dilaton propagator between two points on the cutoff hypersurface ${\cal S}$ is~\cite{Burgess:1984ti}
\begin{equation}
\frac{1}{i} G_\Phi(\tau-\tau',\vec{z}-\vec{z}') = \frac{3}{32\pi^2 R^3} \xi^4 \,_2F_1 (2,5/2 ; 3; \xi^2)  
= \frac{1}{4\pi^2 R^3} \left[ 1 - \frac{1-\frac{3}{2}\xi^2}{(1-\xi^2)^{3/2}} \right]  \ ,
\end{equation}
where $\xi$ is again given by~\ref{xi}.

By the same arguments as above all we need to compute is the Fourier transform $\widetilde{G}_\Phi(\omega,\vec{0})$,
\begin{eqnarray}
\frac{1}{i} \widetilde{G}_\Phi(\omega,\vec{0}) 
 =  \int d\tau d^3z \, e^{i \omega \tau} \frac{1}{i} G_\Phi(\tau,\vec{z})
 =  -\frac{i}{4\pi^2 R^3} \int d\tau_E d^3z \, e^{\omega \tau_E} \left[ 1 - \frac{1-\frac{3}{2}\xi^2}{(1-\xi^2)^{3/2}} \right]           \nonumber  \\
 =  -\frac{i}{4\pi^2 R^3} \int d\tau_E d^3z 
      \left( 1 + \frac{1}{2}\omega^2 \tau_E^2 + \frac{1}{24}\omega^4 \tau_E^4 + \dots \right)
      \left[ 1 - \frac{1-\frac{3}{2}\xi^2}{(1-\xi^2)^{3/2}} \right]   \ .
\end{eqnarray}
Again we have dropped terms that will not contribute to the self-energy of the evaporon after we take the limit $\varepsilon \rightarrow 0$.
Under the change of variables~(\ref{change}) the above expression becomes
\begin{eqnarray}
\widetilde{G}_\Phi(\omega,\vec{0}) 
& = &  \frac{1}{4\pi^2 R^3} {\rm Vol}(S^2) 
    \left( \int_{-\pi/2}^{\pi/2} d\theta \cos^2(\theta) \int_0^\infty d\rho  \, \rho^3 
    \left[ 1 - \frac{1-\frac{3}{2}\xi(\rho)^2}{(1-\xi(\rho)^2)^{3/2}} \right]    \right.         \nonumber \\
& & \quad  +   \frac{\omega^2}{2} \int_{-\pi/2}^{\pi/2} d\theta \cos^2(\theta) \sin^2(\theta) \int_0^\infty d\rho  \, \rho^5 
    \left[ 1 - \frac{1-\frac{3}{2}\xi(\rho)^2}{(1-\xi(\rho)^2)^{3/2}} \right]                    \nonumber \\
& & \quad  \left.  +  \frac{\omega^4}{24} \int_{-\pi/2}^{\pi/2} d\theta \cos^2(\theta) \sin^4(\theta) \int_0^\mu d\rho  \, \rho^7 
    \left[ 1 - \frac{1-\frac{3}{2}\xi(\rho)^2}{(1-\xi(\rho)^2)^{3/2}} \right]    \right)     \ .
\end{eqnarray}
Again, we have introduced an IR cutoff at $\rho = \mu$ in the third term.
Changing integration variable from $\rho$ to $\xi$ we obtain
\begin{eqnarray}
\widetilde{G}_\Phi(\omega,\vec{0}) 
& = &  \frac{\varepsilon^4}{R^3} 
    \int_0^1 d\xi  \, \frac{1-\xi}{\xi^3}  \left[ 1 - \frac{1-\frac{3}{2}\xi^2}{(1-\xi^2)^{3/2}} \right]    \nonumber \\
& &   +  \frac{\omega^2\varepsilon^6}{4 R^3} 
    \int_0^1 d\xi  \, \frac{(1-\xi)^2}{\xi^4}  \left[ 1 - \frac{1-\frac{3}{2}\xi^2}{(1-\xi^2)^{3/2}} \right]    \nonumber \\
& &   +  \frac{\omega^4\varepsilon^8}{48 R^3} 
    \int_{\left(1+\frac{\mu^2}{2\varepsilon^2}\right)^{-1}}^1  d\xi  \, \frac{(1-\xi)^3}{\xi^5}  \left[ 1 - \frac{1-\frac{3}{2}\xi^2}{(1-\xi^2)^{3/2}} \right]    + \dots     \ .
\end{eqnarray}
The integrals over $\xi$ can be done analytically.
The result is
\begin{equation}
\widetilde{G}_\Phi(\omega,\vec{0})  =  \frac{\varepsilon^4}{4 R^3}  +  \frac{\omega^2\varepsilon^6}{24 R^3}  - \frac{\omega^4\varepsilon^8}{128 R^3} \ln\left(\frac{\varepsilon^2}{\mu^2}\right)  +  O(\varepsilon^9)    \ .
\end{equation}
Following the usual procedure we have rescaled $\mu$ to absorb irrelevant numerical constants.

Therefore, the contribution to the evaporon self-energy from the first diagram in Figure~\ref{counterterms} is equal to
\begin{equation}
\left( i \sqrt{2} \kappa \lambda \frac{R^4}{\varepsilon^4} \right)^2 V_3 \frac{1}{i} \widetilde{G}_\Phi(\omega,\vec{0}) 
=  2 i \kappa^2 \lambda^2 R^5 V_3  \left[ \frac{1}{4 \varepsilon^4}  +  \frac{\omega^2}{24 \varepsilon^2}  -  \frac{\omega^4}{128} \ln\left(\frac{\varepsilon^2}{\mu^2}\right)  +  O(\varepsilon)  \right]     \ .
\end{equation}
Cancellation of the divergences then uniquely specifies the counterterms that must be added to the action~(\ref{action}):
\begin{equation}
S_{c.t.}[\sigma] =  \kappa^2 \lambda^2 R^5 V_3  \int d\tau dx \, \delta(x) 
  \left[ - \frac{1}{4 \varepsilon^4} \sigma^2  -  \frac{1}{24 \varepsilon^2} \dot{\sigma}^2
  +  \frac{1}{128} \ln\left(\frac{\varepsilon^2}{\mu^2}\right) \ddot{\sigma}^2   \right]     \ .
\label{counter2}
\end{equation}

\section{Scattering off the interface}

We now consider an outgoing wave for the $\Phi$ field reaching the boundary of AdS and partially reflecting, with the remaining energy being transmitted into the extra space in the form of an outgoing evaporon wave.
We will determine the transmission coefficient in this way but we can also consider an incoming evaporon partially reflecting.
Both calculations are expected to yield the same result.

The equations of motion can now be derived from the action~(\ref{action}) together with equations~(\ref{counter1}) and~(\ref{counter2}):
\begin{eqnarray}
y^2\partial_y^2\Phi &-& 3y\partial_y\Phi - y^2\partial_\tau^2\Phi - \Delta(\Delta - 4)\Phi
=  - 2 \kappa^2 \lambda \, R \, \varepsilon^{5-\Delta} \, \delta(y-\varepsilon) \, \sigma(\tau,0)  \ , \label{dilatonEOM} \\ 
\nonumber \\
-\partial_\tau^2\sigma &+& \partial_x^2\sigma
=  - \delta(x) \lambda V_3 \, R^4 \varepsilon^{-\Delta} \Phi(\tau,\varepsilon) + \delta(x) \, 2\kappa^2 \lambda^2 R^5 V_3 \, f(\sigma)     \ ,
\label{evaporonEOM}
\end{eqnarray}
where
\begin{eqnarray}
f(\sigma) & = &  - \frac{\sigma}{2}  \ln\left(\frac{\varepsilon^2}{\mu^2}\right)  \hspace{1.3 in} {\rm for} \: \Delta = 2  \ , \\
f(\sigma) & = &  \frac{\sigma}{4\varepsilon^4} - \frac{\ddot{\sigma}}{24\varepsilon^2}  - \frac{\sigma^{(4)}}{128} \ln\left(\frac{\varepsilon^2}{\mu^2}\right)    \quad\quad {\rm for} \: \Delta = 4  \ ,  
\end{eqnarray}
and we are implicitly assuming the s-partial wave for the bulk field so that it is independent of the transverse coordinates $\vec{z}$.

In the bulk region ($y>\varepsilon$) scalar field modes with definite frequency $\omega$ can be expressed as a linear combination of two independent solutions of eq.~(\ref{dilatonEOM}) :
\begin{equation}
\Phi(\tau,y) = e^{- i \omega \tau} \left[ \beta \, y^2 J_{\Delta-2}(\omega y) + \gamma \, y^2 Y_{\Delta-2}(\omega y) \right]  +  h.c.   \ ,
\label{dilatonBULK}
\end{equation}
whereas in the boundary region ($y<\varepsilon$) normalizability of the field requires one of the coefficients to vanish and so
\begin{equation}
\Phi(\tau,y) = \alpha \, e^{- i \omega \tau} y^2 J_{\Delta-2}(\omega y)  +  h.c.    \ .
\label{dilatonBDY}
\end{equation}

For the evaporon the solutions are simply incoming and outgoing plane waves.
However, we are interested in the case in which there is no incoming wave.
Thus, we take it to be of the form
\begin{equation}
\sigma(\tau,x) = B e^{- i \omega (\tau - x)}  +  h.c.   \ ,
\label{evaporonSOL}
\end{equation}
for $x>0$.
Recall from section~2 that we require $\sigma$ to be an even function in the spatial coordinate, $\sigma(\tau,x)= \sigma(\tau,-x)$.

Equivalently, we could also consider an incident evaporon wave on the interface which is partially reflected while the remaining fraction of energy is transmitted into an ingoing dilaton wave in ${\rm AdS}$.
Requiring the solution for the bulk field to be ingoing is equivalent to setting $\gamma = i \beta$.
In this scenario the evaporon takes the form
\begin{equation}
\sigma(\tau,x) = A e^{-i \omega (\tau + x)} + h.c.  + B e^{-i \omega (\tau - x)} + h.c.  \ .
\end{equation}

Continuity of the solution for the bulk field is implemented by
\begin{equation}
(\beta - \alpha) \, J_{\Delta-2}(\omega \varepsilon) + \gamma \, Y_{\Delta-2}(\omega \varepsilon) = 0   \ ,
\label{contin}
\end{equation}
and as usual, the $\delta$-functions on the right-hand side of equations~(\ref{dilatonEOM}) and~(\ref{evaporonEOM}) result in discontinuities of the first derivatives:
\begin{eqnarray}
{\rm Disc} \left. \partial_y \Phi \right|_{y=\varepsilon} &=& - 2 \kappa^2 \lambda R \varepsilon^{3-\Delta} \sigma(\tau,0) \ ,
\label{dilatonDISC} \\
{\rm Disc} \left. \partial_x \sigma \right|_{x=0}  &=&  - \lambda V_3 \, R^4 \varepsilon^{-\Delta} \Phi(\tau,\varepsilon) 
  + 2 \kappa^2 \lambda^2 R^5 V_3 \, f(\sigma(\tau,0))    \ .
\label{evaporonDISC}
\end{eqnarray}
These conditions on the discontinuities of the first derivatives together with the condition~(\ref{contin}) are sufficient to solve for the scattering coefficients $\alpha$, $\beta$ and $\gamma$ in terms of the remaining coefficient $B$.
But before we do so let us pause to obtain an expression for the transmission coefficient.

\section{The transmission coefficient}

In this section we calculate the transmission coefficient in terms of the parameters $\beta$ and $\gamma$.
We will work explicitly with the $\Delta = 4$ case but the final result is the same for $\Delta = 2$.

Imagine a dilaton wave incident from the bulk of ${\rm AdS}_5$ and reflecting back at the hypersurface $y=\varepsilon$.
Using the behavior of the Bessel functions for large argument ($r \gg 1$):
\begin{eqnarray}
J_{\nu}(r) & \sim & \sqrt{\frac{2}{\pi r}} \cos \left( r-\nu\frac{\pi}{2}-\frac{\pi}{4} \right) \ , \\
Y_{\nu}(r) & \sim & \sqrt{\frac{2}{\pi r}} \sin \left( r-\nu\frac{\pi}{2}-\frac{\pi}{4} \right) \ ,
\end{eqnarray}
we see that for large $y$ the dilaton behaves like $\Phi_{inc} + \Phi_{ref}$, where the incident part is given by
\begin{equation}
\Phi_{inc}(\tau,y) = - \frac{y^{3/2}}{\sqrt{2\pi \omega}} (\beta+i\gamma) e^{i\pi/4} e^{-i\omega(\tau+y)} + h.c. \ ,
\end{equation}
and the reflected part is
\begin{equation}
\Phi_{ref}(\tau,y) = - \frac{y^{3/2}}{\sqrt{2\pi \omega}} (\beta-i\gamma) e^{-i\pi/4} e^{-i\omega(\tau-y)} + h.c. \ ,
\end{equation}

To obtain the reflection coefficient we consider the time average of the energy density coming from the incident wave and from the reflected wave using the action~(\ref{action}).
These are given respectively by
\begin{eqnarray}
\left< \rho_{inc} \right>(y) = \frac{9}{8\pi\omega y^2} \left[ |\beta|^2 + |\gamma|^2 + 2\,{\rm Im}(\beta\gamma^*) \right] \ , \\
\left< \rho_{ref} \right>(y) = \frac{9}{8\pi\omega y^2} \left[ |\beta|^2 + |\gamma|^2 - 2\,{\rm Im}(\beta\gamma^*) \right] \ ,
\end{eqnarray}
so that the reflection coefficient is
\begin{equation}
|{\cal R}|^2 \equiv \frac{\left<\rho_{ref}\right>}{\left<\rho_{inc}\right>} 
= 1 - \frac{4\,{\rm Im}(\beta\gamma^*)}{|\beta|^2 + |\gamma|^2 + 2\,{\rm Im}(\beta\gamma^*)} \ .
\label{refcoef}
\end{equation}

Thus, we conclude that we need the coefficients $\beta$ and $\gamma$ to scale with $\varepsilon$ in a similar fashion in order to obtain partial transmission of the waves at the boundary after we take the limit $\varepsilon \rightarrow 0$.
The coefficients $\beta$ and $\gamma$ derived from the action~(\ref{action}) ignoring the $S_{c.t.}$ piece do not scale accordingly.
However, this is precisely remedied by the addition of the counterterms we have found in section~3.
For the situation of an incident evaporon wave on the interface the reflection coefficient takes the expected form:
\begin{equation}
|{\cal R}|^2 = \frac{|B|^2}{|A|^2}  \ .
\label{refcoef2}
\end{equation}

We are now able to complete the scattering calculation and obtain the reflection and transmission coefficients.
Using the (dis)continuities relations~(\ref{contin} -- \ref{evaporonDISC}) we can express $\beta$ and $\gamma$ in terms of the coefficient $B$:
\begin{eqnarray}
\gamma &=& B \frac{2 \lambda \kappa^2 R}{\omega} \, \frac{\varepsilon^{-3} J_2(\omega \varepsilon)}
    {J_1(\omega \varepsilon) Y_2(\omega \varepsilon) - J_2(\omega \varepsilon) Y_1(\omega \varepsilon)}    \ , 
  \\
\beta &=& B \frac{2 \lambda \kappa^2 R}{\omega} \, \frac{\varepsilon^{-3} Y_2(\omega \varepsilon)}
    {J_2(\omega \varepsilon) Y_1(\omega \varepsilon) - J_1(\omega \varepsilon) Y_2(\omega \varepsilon)}  
  \nonumber \\
     && + B \frac{\varepsilon^2}{J_2(\omega \varepsilon)} \left[ \frac{2 i \omega}{\lambda R^4 V_3} 
       - \frac{\lambda \kappa^2 R}{2}  \left\{ \frac{1}{\varepsilon^4} + \frac{\omega^2}{6\varepsilon^2} 
       - \frac{\omega^4}{32} \ln \frac{\varepsilon^2}{\mu^2}  \right\} \right]   \ .
\end{eqnarray}
These expressions are valid for the $\Delta=4$ case.
Now, if we take the limit $\varepsilon \rightarrow 0$ both coefficients remain finite:
\begin{eqnarray}
\gamma & \rightarrow &  - \frac{\pi}{8} \lambda \kappa^2 \omega^2 R \, B   \ , 
  \\
\beta & \rightarrow &  - \frac{16 i}{\lambda R^4 V_3 \omega} B + \frac{\lambda \kappa^2 R \omega^2}{4} B \ln (\mu \, \omega) 
\ .
\end{eqnarray}
Therefore, using eq.~(\ref{refcoef}) we obtain the following formula for the transmission coefficient:
\begin{equation}
|{\cal T}|^2 = 1 - |{\cal R}|^2  =  \frac{2}{1 + \frac{1}{4\pi}\left(\frac{\omega_4}{\omega}\right)^3 + \pi\left(\frac{\omega}{\omega_4}\right)^3 \left[ 1 + \frac{4}{\pi^2} (\ln (\mu \, \omega) )^2 \right]} \ ,
\label{trans}
\end{equation}
where we have defined $\omega_4 \equiv  8 (2 \lambda^2 \kappa^2 R^5 V_3)^{-1/3}$.
The result (see Figure~\ref{transmission}) depends logarithmically on the scale $\mu$ but all the dependence on the cutoff $\varepsilon$ has canceled out as necessary.
Similarly, we could have solved the problem of an incident evaporon wave partially transmitting through the interface.
Using eq.~(\ref{refcoef2}) this leads to the same result.

\FIGURE[t]{
 \includegraphics[width=25pc, bb= 0pt 0pt 395pt 255pt]{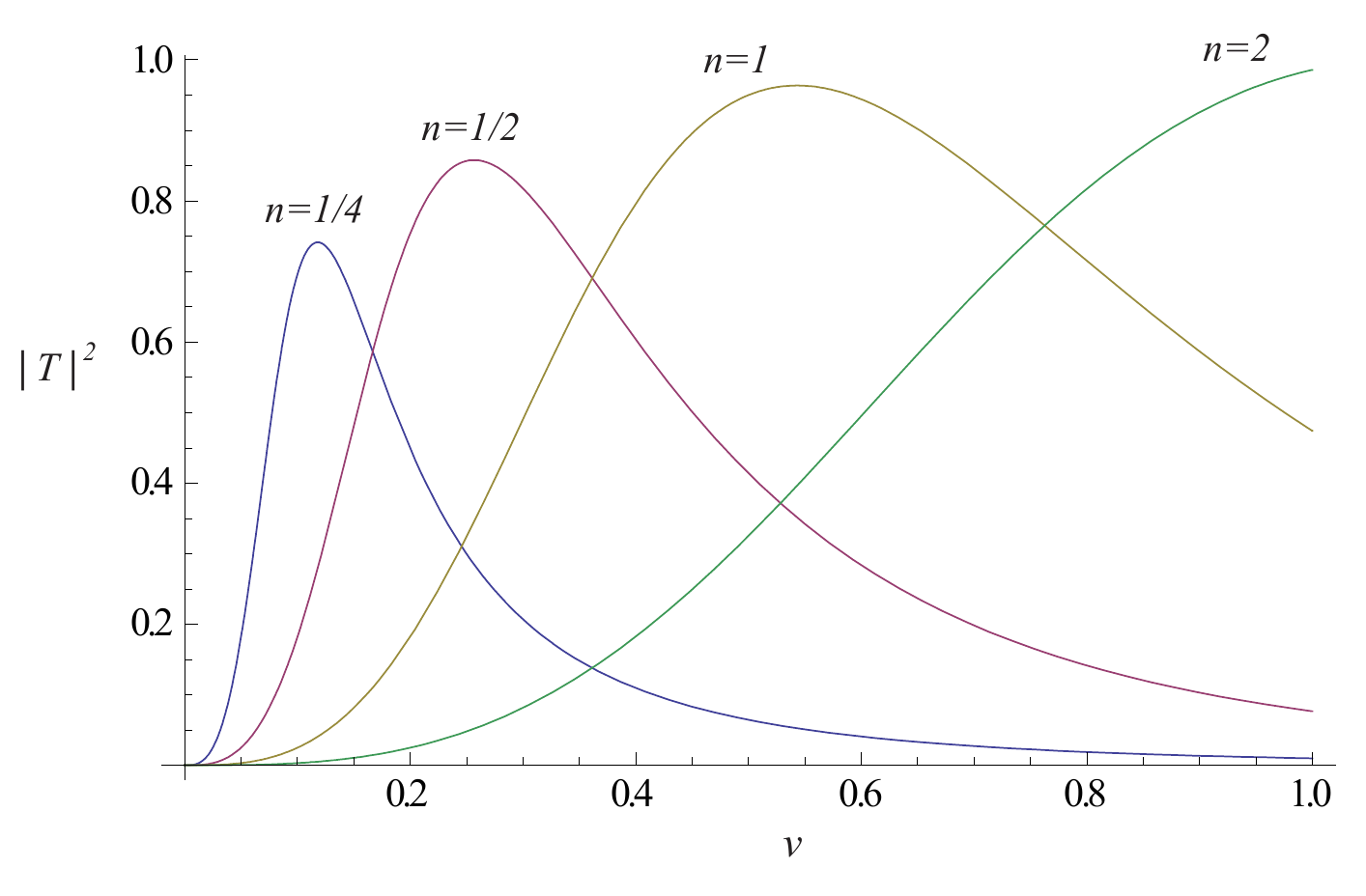} 
 \caption{The transmission coefficient is plotted as a function of the frequency in units of the UV cutoff scale $\nu \equiv \mu\,\omega$.  The result is shown for four different values of the effective coupling constant, equally in units of the UV cutoff scale, $n \equiv \mu\,\omega_4$.}
 \label{transmission}
}

The result for the $\Delta=2$ case is very similar.
While the coefficients $\gamma$ and $\beta$ are given by
\begin{eqnarray}
\gamma & \rightarrow &  - \pi \lambda \kappa^2 R \, B   \ , 
  \\
\beta & \rightarrow &  - \frac{2 i \omega}{\lambda R^4 V_3} B + 2 \lambda \kappa^2 R B \ln (\mu \, \omega) 
\ ,
\end{eqnarray}
the transmission coefficient is almost identical to~(\ref{trans}).
Indeed, we can package the two cases into a single formula:
\begin{equation}
|{\cal T}|^2 =  \frac{2}{1 + \frac{1}{4\pi}\left(\frac{\omega_\Delta}{\omega}\right)^{2\Delta-5} + \pi\left(\frac{\omega}{\omega_\Delta}\right)^{2\Delta-5} \left[ 1 + \frac{4}{\pi^2} (\ln (\mu \, \omega) )^2 \right]} \ ,
\label{transmit}
\end{equation}
with $\omega_2 \equiv \lambda^2 \kappa^2 R^5 V_3 / 4$.

Figure~\ref{transmission} shows that the transmission coefficient has two branches.
For low frequencies it is an increasing function but then it turns around and becomes decreasing.
As a result, for each value of the coupling constant $\lambda$ (or equivalently of $\omega_\Delta$) there is an optimal frequency to transfer energy out of the bulk.
This can be understood as follows.
If we discard the logarithmic correction in equation~(\ref{transmit}) we can trade the dependence on the frequency by a dependence on the coupling.
When the coupling is weak only a small fraction of the wave in the bulk is absorbed at the interaction interface and the rest gets reflected at the boundary of AdS.
On the opposite extreme, when the coupling is strong, most of the wave does not even reach the boundary and instead bounces off the interface.
Interestingly, we have found that which of the two coupling regimes (weak or strong) corresponds to low energy scattering depends on the conformal dimension $\Delta$.
The meaning of this curious fact is not clear at the moment.

\section{Conclusions and discussion}

In conclusion, we have presented a toy model that allows large black holes in AdS to evaporate.
This was achieved by coupling, at the boundary of AdS, a bulk scalar field representing the Hawking radiation to an external scalar field.
Such a modification effectively changes the boundary conditions so that it becomes only partially reflective, permitting some energy to leak out of AdS.
In the dual gauge theory description this situation corresponds to adding a weakly coupled sector to the strongly coupled CFT.
The large black hole represents a high temperature thermal state and the evaporation of the former is associated with the cool down of the CFT by transferring energy to the external sector.

We have computed the transmission coefficient in the framework of the toy model.
This was done perturbatively to second order in the coupling constant and involved careful regularization and addition of appropriate counterterms.
We found a resonant-like behavior for the transmission coefficient as a function of the frequency: bulk modes only decay efficiently if they have a frequency close to the resonance.

One could also be concerned about the evaporon mass that enters through the counterterms back-reacting on the geometry.
This would be incorporated by inclusion of interactions with gravitons but in the large $N$ limit we are considering we can safely ignore this effect.

It would be interesting to compute the rate of evaporation allowed by this scenario for the large black holes in an asymptotically anti-de Sitter spacetime.
In principle, this can be done along the lines of~\cite{Hawking:1974sw}.
The calculation essentially amounts to finding the overlaps between the evaporon modes in the past and future null infinities.
The result obtained above for the transmission coefficient will then enter the evaporation rate in the same way as a gray-body factor, i.e. it is a multiplicative factor that distorts the spectrum of the Hawking radiation.
We hope to report on this work in the near future.

A full account of the evaporation process is beyond the scope of this paper but we see no reason to suspect that the features of our model ruin the evaporation of the black hole, at least while we remain in the class of large black holes in AdS.
Of course, the transmission coefficient relies solely on properties of the boundary of AdS and the coupling to the auxiliary dimension, so it does not depend on the size of the black hole, in contrast with the gray-body factor.
One might worry that, as the black hole shrinks and the peak of the radiation spectrum (with the gray-body factor included) shifts toward lower frequencies, we reach a point where the evaporation halts.
Indeed, even if we start with the ideal situation in which the coupling $\omega_\Delta$ is chosen such that the peak in the transmission coefficient fits the radiation spectrum, a mismatch will develop as the black hole evaporates.
However, the rate of energy loss from the black hole is determined by the overlap of these two {\em finite} functions of the frequency and so there should be no concern about the evaporation process ever stopping.
Stated equivalently, no matter what the characteristic frequency of the radiation reaching the boundary of AdS is, there will always be some fraction of that energy lost to the auxiliary dimension.
Therefore, the rate of shrinkage may decrease but it will not stop.

Usually the information paradox is presented in terms of an evolution from a non-singular pure state before the formation of a black hole into an also non-singular mixed state which is expected to be left after a black hole evaporates.
The black hole formation can be easily accommodated in our model: one can simply imagine a collapsing spherical shell of matter being sent in from the auxiliary flat space and propagating into AdS, consequently producing a black hole.
Only a fraction of the shell will be transmitted at the interface but there is no upper limit on the amount of energy one can initially endow the collapsing shell with. 

As the black hole shrinks it will eventually cease to belong to the class of AdS large black holes and the dual field theoretic description becomes less clear at that point.
Nevertheless, the evaporation process is expected to continue at least until the Hawking-Page transition, when the size of the black hole is comparable to the AdS scale $R$.
Note that one does not need complete evaporation to arrive at the information paradox: according to \cite{Page:1993wv}, one starts recovering information encoded in the Hawking radiation after half of the initial entropy of the black whole has been radiated, which contradicts the semi-classical result of a black-body spectrum for the radiation.
One may then simply consider forming a black hole with Schwarzschild radius $r_S \gg R$ and letting it evaporate until $r_S \sim R$ so that a version of the information paradox still applies.

\acknowledgments

I wish to thank J. Polchinski for suggesting this project, for many enlightening conversations along the road and for comments on a draft.  I also thank G. Horowitz and J. Penedones for useful discussions.  This work was supported by NSF grant PHY04-56556.


\end{document}